\documentclass[
  aps,
  pra,
  reprint,
  10pt,
  tightenlines,
  showpacs,
  nofootinbib,
  superscriptaddress
]{revtex4-2}

\usepackage{amsmath,amssymb,graphicx}
\usepackage{physics}
\usepackage{multirow}
\usepackage[normalem]{ulem}
\usepackage{braket}
\usepackage{bm}
\usepackage{balance}
\usepackage{xcolor}
\usepackage{tikz}
\usepackage{quantikz}
\usepackage[caption=false]{subfig}
\usepackage{hyperref} 

\usepackage{hyperref}

\usepackage[justification=raggedright,singlelinecheck=false]{caption}

\newcommand{\blk}{\color{black}}

\begin{document}

\title{Optimal and robust error filtration for quantum information processing}

\author{Aaqib Ali}
\affiliation{Dipartimento Interateneo di Fisica, Università di Bari, 70126 Bari, Italy}
\affiliation{INFN, Sezione di Bari, 70126 Bari, Italy}

\author{Giovanni Scala}
\affiliation{Dipartimento Interateneo di Fisica, Politecnico di Bari, 70126 Bari, Italy}
\affiliation{INFN, Sezione di Bari, 70126 Bari, Italy}

\author{Cosmo Lupo}
\affiliation{Dipartimento Interateneo di Fisica, Politecnico di Bari, 70126 Bari, Italy}\affiliation{Dipartimento Interateneo di Fisica, Università di Bari, 70126 Bari, Italy}
\affiliation{INFN, Sezione di Bari, 70126 Bari, Italy}

\begin{abstract}

Error filtration is a hardware scheme that mitigates noise by exploiting auxiliary qubits and entangling gates. Although both signal and ancillas are subject to local noise, constructive interference and post-selection allow us to reduce the noise level in the signal qubit.
Here we highlight the relation between error filtration and error detection codes, and determine the optimal codes that make the qubits interfere most effectively.
We examine our optimized scheme under imperfect implementation, where ancillary qubits may be noisy or subject to cross-talk. 
Even with these imperfections, we find that adding more ancillary qubits helps in protecting quantum information.
We benchmark our approach against figures of merit that correspond to different applications, including entanglement fidelity, quantum Fisher information (for applications in quantum sensing), and CHSH value (for cryptographic applications), with one, two, and three ancillary qubits. 
By using the entanglement fidelity as a figure of merit, we suggest a general condition for error filtration 
and, for one and two ancillas, we obtain some explicit expressions for the optimal codes.
We also compare our method with the recently introduced Superposed Quantum Error Mitigation (SQEM) scheme based on superposition of causal orders, and show that, for a wide range of noise strengths, our approach may outperform SQEM in terms of effectiveness and robustness.
\end{abstract}

\maketitle

\section{Introduction} \label{sec:introduction}

Quantum information science has opened the door to revolutionary advancements in computing~\cite{degen2017quantum,Dowling,JPD}, sensing~\cite{PhysRevLett.109.070503,pepe2022distance,Tsang,Scala2024c}, and communication~\cite{Primaatmaja2023,Zapatero2023}, but one of the biggest hurdles remains maintaining coherence. As we navigate the Noisy Intermediate-Scale Quantum (NISQ) era~\cite{preskill2018quantum}, suppressing and mitigating the effects of noise and errors is critical~\cite{endo2018practical,strikis2021learning,cai2023quantum,endo2021hybrid,suzuki2022quantum,acampora2021genetic,lostaglio2021error}. 
In this work, we develop schemes of \textit{error filtration}, a method introduced by Gisin \textit{et al.}~\cite{PhysRevA.72.012338} to filter noise via superposition and post-selection.

\begin{figure}[t]
\resizebox{0.9\linewidth}{!}{
\begin{quantikz}[row sep={0.8cm,between origins}, column sep=0.2cm]
     \lstick[1]{$\ket{\psi}$} & \qw & \gate[wires=2,style={fill=green!20}]{\, U \, } \slice{} & \qw  & \gate[style={fill=red!20}]{ \hspace{0.2cm} \mathcal{E} \hspace{0.2cm} } & \qw \slice{} & \gate[wires=2,style={fill=green!20}]{U^\dagger} & \qw  \\
     \lstick[1]{$\ket{0}^{\otimes n}$} &  \qwbundle{n} & \qw & \qw & \gate[style={fill=red!20}]{\mathcal{E}^{\otimes n}} & \qw & \qw & \meter{}\\
     & \wireoverride{}&\wireoverride{} \text{Encoding} & \wireoverride{} &\wireoverride{} & \wireoverride{} &\wireoverride{}\text{Decoding}
\end{quantikz}
}
\caption{
General scheme of error filtration. 
On the left, one signal qubit $|\psi\rangle$, and $n$ ancillary qubits $|0\rangle^{\otimes n}$. 
After the encoding $U$ on the signal and ancillas, the information is transmitted through a memoryless noisy channel $\mathcal{E}$.
The inverse unitary $U^\dag$ is applied for the decoding,
followed by a measurement of the ancillary qubits. 
The effects of noise on the signal qubit are reduced conditioned on the ancillary qubits being found in the state 
$|0\rangle^{\otimes n}$.
}
\label{circuit}
\end{figure}

The general scheme of error filtration is shown in Fig.~\ref{circuit}. An input state $\ket{\psi}$ is subject to a memoryless noisy channel, denoted $\mathcal{E}$, which degrades information and introduce errors  ---  modeling e.g.~storage in a quantum memory or transmission through an optical fiber.
To protect the quantum information in $\ket{\psi}$, error filtration exploits $n$ auxiliary systems initially prepared in $\ket{0}^{\otimes n}$, which are coupled to the signal via the encoding unitary $U$.
When retrieving the information, the decoding process is performed by applying the inverse unitary $U^\dag$ followed by post-selection on the ancillas. For suitable choice of the unitary $U$, an increase in the fidelity is observed conditioned on the ancillas being found in the state $\ket{0}^{\otimes n}$.
In contrast to deterministic error correction, error filtration is in general probabilistic. 
However, as we show below, the post-selected state may have higher fidelity than the deterministic outcome of error correction. This is especially true when the number of ancillas is limited, whereas for sufficiently many ancillary systems one expects quantum error correction to be optimal.

The challenge we address here is to determine the \textit{optimal} unitary acting on
$N=n+1$ qubits given a noise model, 
a task that is computationally demanding due to the exponential growth of the $2^{2N}-1$ parameters in $\text{SU}(2^{N})$.
Nevertheless, it is possible to find the optimal encoding unitary that protects information despite the noise and cross-talk acting on the signal and ancillary qubits.
We show that an error filtration scheme can be obtained from an error \textit{detection} code. 
We suggest a general condition for error filtration, for any noise model, that resembles the condition of error detection~\cite{KnillLaflamme}. We provide examples of unitary encoding that fulfil such a condition and are found to be optimal.

Our filtration scheme is optimized according to specific figures of merit, determined by the particular goal achieved through quantum information processing.
For computation and general quantum information processing, we focus on the entanglement fidelity~\cite{mandarino2016assessing,nielsen2010quantum,mandarino2014use}. 
The CHSH functional is considered as a theoretical tool to study non-locality and for implementing device-independent protocols~\cite{Brunner2014,Scarani2019,Xu2020,Portmann2022,Gigena2024}.
Finally, for metrology and sensing applications we examined the quantum Fisher information (QFI)~\cite{Sidhu}.
This latter case is peculiar as, for our model, error filtration and error correction yield equivalent results.

Unlike other works
based on variational methods, see e.g.~\cite{Lukin2019}, 
or focused on specific subclasses of encoding unitaries~\cite{PhysRevLett.131.230601,PhysRevA.108.062604,lee2023error,vijayan2020robust,huang2023,karthik2024noise}, 
our optimization scheme, although 
computationally demanding, is ansatz-free as we aim at finding the globally optimal encoding and decoding unitaries.
Remarkably, through optimization over universal $U\in \mathrm{SU}(2^N)$ gates, our work generalizes and encompasses previous results~\cite{lee2023error, PhysRevLett.131.230601, PhysRevA.108.062604} that focused on specific encodings.
We investigate systems with up to three auxiliary qubits, providing practical solutions for error reduction in photonic (e.g.~for adapting multicore fibers to quantum communication) and for the characterization of quantum memories~\cite{Stein2024,PerezCastro2025}, each corresponding to a different figure of merit~\cite{RojasRojas2024nonmarkovianityin,Martinez2022,Rojas21}.

The paper develops as follows. 
In Section~\ref{sec:univ} we introduce a suitable parameterization for the encoding and decoding unitary, and the noise model.
Numerical and analytical results are presented in Section~\ref{sec:results}. 
In particular, the robustness of error filtration under imperfect implementation is discussed in Section~\ref{sec:noiseenc}.
A comparison with the SQEM approach of Ref.~\cite{PhysRevLett.131.230601} is presented in Section~\ref{sec:sqem}. Conclusions and an overview of future developments are presented in Section~\ref{sec:end}.

\section{Error filtration from universal unitary gates}\label{sec:univ}

In this Section, we consider an error-free unitary $ U $ (see Fig.~\ref{circuit}) to encode the signal qubit
with the support of $n$ auxiliary qubits initially prepared in $\ket{0}$. 
All the $N=n+1$ qubits are sent through a memoryless noisy channel 
$\mathcal{E}$ as described in Sec.~\ref{subsec:noise}.

\begin{figure}[t]
\centering
\resizebox{0.45\textwidth}{!}{
\begin{quantikz}
    \lstick{} & 
    \gate{U_1} & \targ{}  & \gate{R_Z} &\ctrl{1}& \qw & \targ{} &\gate{U_3} & \qw \\
    \lstick{} 
    & \gate{U_2} & \ctrl{-1}  & \gate{R_Y} &\targ{}&\gate{R_Y} &\ctrl{-1}& \gate{U_4} & \qw
\end{quantikz}}
\caption{
Construction of a two-qubit unitary gate from four single-qubit gates $U_1, U_2, U_3, U_4 \in \mathrm{SU}(2)$, three CNOT gates, two rotations along the $y$-axis, and one along the $z$-axis~\cite{shende2004minimal}.
}
\label{fig:U-2-qubits}
\end{figure}

\subsection{Universal unitary gates}\label{subsec:gates}

We use universal unitary gates for their ability to approximate any quantum operation, enabling a wide range of applications.

The literature offers multiple decomposition strategies, including Givens rotations and sine-cosine decompositions, which provide a structured methodology for simplifying complex quantum gates into more manageable subunits (see Refs.~\cite{mottonen12006decompositions,khan2005synthesis,krol2022efficient,barenco1995elementary,bullock2003elementary,cybenko2001reducing,mottonen2004quantum,paige1994history,vartiainen2004efficient,Wiersema2024}).

For the specific task of constructing a two-qubit unitary transformation (i.e.~for an error filtration scheme with $n=1$ ancillary qubit), we adopt the circuit architecture described in Ref.~\cite{shende2004minimal}. 
Figure~\ref{fig:U-2-qubits} illustrates how a generic two-qubit unitary gate $U \in \mathrm{U}(2^2)$ can be synthesized from single-qubit gates and the two-qubit gate CNOT, where 
$R_X = e^{-i\theta\sigma_x/2}$, 
$R_Y = e^{-i\phi\sigma_y/2}$, and 
$R_Z = e^{-i\gamma\sigma_z/2}$, where $\sigma_x$, $\sigma_y$, $\sigma_z$ are the Pauli matrices.
(These can be combined following Euler decomposition to obtain general one-qubit unitaries.)

To expand this construction to more qubits (number of ancillas $n \geq 2$), we employ the Quantum Shannon Decomposition (QSD) described in Ref.~\cite{shende2005}
{(other methods, such as the Cartan KAK decomposition~\cite{tucci2005introduction}, yield equivalent results).}
QSD recursively allows for the implementation of an arbitrary $(n+1)$-qubit unitary operation through a sequence of multiplexed rotations and $n$-qubit unitaries,
arranged in a specific circuit configuration as depicted in Figs.~\ref{fig:QDSdecomposition}-\ref{fig:nesting2}.
However, while the QSD recursion is adaptable to circuits with any number of qubits, it does increase the number of required C-NOT gates. This trade-off is managed by recursively applying the decomposition until the circuit simplifies to two-qubit gates, after which we use the minimal decomposition~\cite{shende2005}.

The QSD depends on 15, 72, and 312 parameters, respectively for $n=1,2,3$ ancillas. 
Note that these numbers exceed $2^{2N}-1$ free parameters of $\text{SU}(2^N)$. This is the cost we pay to have a decomposition that is more suitable for experimental implementations.

\begin{figure}
\centering
\resizebox{0.45\textwidth}{!}{
\begin{quantikz}
&\qwbundle{n} & \gate{V_1} & \octrl{1} & \gate{V_2} & \octrl{1} & \gate{V_3} & \octrl{1} & \gate{V_4} & \qw \\
& \qw & \qw & \gate{R_Z} & \qw & \gate{R_Y} & \qw & \gate{R_Z} & \qw & \qw 
\end{quantikz}}
\caption{
Quantum Shannon Decomposition for constructing \(U\in\mathrm{U}(2^{n+1})\) involving recursive application of \(V_1, V_2, V_3, V_4\in\mathrm{U}(2^{n})\) and multiplexed rotations~\cite{shende2005}.
The recursive definition of the multiplexed rotation is shown in Fig.~\ref{fig:nesting}.
}
\label{fig:QDSdecomposition}
\end{figure}


\begin{figure}
\centering
\resizebox{0.45\textwidth}{!}{
\begin{quantikz}
 & \qwbundle{n} & \octrl{1} & \qw & \qw  \\
 & \qw & \gate{R_Z} & \qw & \qw      
\end{quantikz} 
\quad \text{\huge $=$} 
\begin{quantikz}[thin lines]
   & \qw & \qw & \ctrl{2} & \qw & \ctrl{2} & \qw  \\
    &\qwbundle{n-1} &\octrl{1} & \qw& \octrl{1} & \qw & \qw  \\
   & \qw & \gate{R_Z} & \targ{} & \gate{R_Z} & \targ{} & \qw
\end{quantikz}
}
\caption 
{By recursively applying the decomposition shown in the picture, one can decompose any multiplexed rotation into elementary gates~\cite{shende2005}.}
\label{fig:nesting}
\end{figure}

\begin{figure}[htb!]
\centering
\resizebox{0.45\textwidth}{!}{
\begin{quantikz}
 & \qw & \octrl{1} & \qw & \qw  \\
 & \qw & \gate{R_Z} & \qw & \qw      
\end{quantikz} 
\quad \text{\huge $=$} 
\begin{quantikz}[thin lines]
   & \qw & \qw & \ctrl{1} & \qw & \ctrl{1} & \qw  \\
   & \qw & \gate{R_Z} & \targ{} & \gate{R_Z} & \targ{} & \qw
\end{quantikz}
}
\caption{For \( n=1 \), the multiplexed rotation is defined by combining two C-NOT gates and two single-qubit rotations. Note that in general the two single-qubit rotations have different angles~\cite{shende2005}.
}
\label{fig:nesting2}
\end{figure}

\subsection{Noise models}\label{subsec:noise}

In the error filtration circuit depicted in Fig.~\ref{circuit} we assume a consistent noisy channel affecting both signal and ancillary qubits. 
As an example, we explore two prevalent models of quantum noise: the \textit{dephasing channel} and the \textit{depolarizing channel}. In the following, we recall their action in a Bell state $\ket{\psi}$.
The dephasing channel, denoted as $\mathcal{E}_\varphi$, affects the relative phase between $\ket{0}$ and $\ket{1}$ in the following Kraus representation:
\begin{align}
\{\sqrt{p_\varphi} \, \sigma_0 \, , 
\sqrt{1-p_\varphi} \, \sigma_z\} \, ,
\label{eq:dephasing_noise}
\end{align}
where $\sigma_0$ denotes the identity matrix.
The action of this channel on the Bell state preserves the states in the computational basis states but introduces phase errors:
\begin{equation}
\mathcal{E}_\varphi\left( |\psi\rangle \langle \psi| \right) = 
q_\varphi
|\psi\rangle \langle \psi|
+ 
(1-q_\varphi) \tau_\psi \, ,
\label{eq:state_dephase_noise}
\end{equation}
where $q_\varphi = 2 p_\varphi - 1$ and
\begin{align}
    \tau_\psi = \frac{|\psi\rangle \langle \psi|
    + \sigma_z |\psi\rangle \langle \psi| \sigma_z}{2}
\end{align}
is the state with erased phase information. The depolarizing channel, denoted as $\mathcal{E}_r$, models the loss of quantum information without bias towards any specific basis. Its effect on a quantum state can be described using the Kraus operators:
\begin{align}
\left\{
\sqrt{p_r} \, \sigma_0 \, , 
\sqrt{\frac{1-p_r}{3}} \, \sigma_x \, ,
\sqrt{\frac{1-p_r}{3}} \, \sigma_y \, ,
\sqrt{\frac{1-p_r}{3}} \, \sigma_z\right\} \, .
\label{eq:depolarizing_noise}
\end{align}
The channel transforms any input state towards the maximally mixed state $\sigma_0/2$. 
The action on a pure state $|\psi\rangle$ is
\begin{align}
    \mathcal{E}_r(|\psi\rangle \langle \psi|) 
    =
    q_r |\psi\rangle \langle \psi|
    + (1-q_r) \, \frac{\sigma_0}{2} \, ,
    \label{state_depolarizing_noise}
\end{align}
where the parameter $q_r = (4 p_r -1)/3$ quantifies the probability that the input state remains unchanged, and $(1-q_r)$ is the probability that it is replaced by the maximally mixed state.

\section{Numerical results and theoretical insights}\label{sec:results}

In this Section, we denote as $\mathfrak{F}$ a generic cost function, such as entanglement fidelity, the CHSH functional, or quantum Fisher information, which acts on the normalized state
$\rho^{\mathrm{out}}_U$.
We remark the dependence of the state on the unitary $U\in \mathrm{SU}(2^{n+1})$ that encodes the initial state made of an input signal qubit $\ket{\psi}$ and ancillary qubits initialized in $\ket{0}^{\otimes n}$. The optimal value of $\mathfrak{F}_n^*$ is given by
\begin{equation}
\mathfrak{F}_n^* = \max_{U \in \mathrm{SU}(2^{n+1})} \mathfrak{F}[\rho^{\mathrm{out}}_U] \, .
\end{equation}
As the number of ancillae increases, the functional improves since $\mathrm{SU}(2^{n+1})$ is a subgroup of $\mathrm{SU}(2^{n+2})$, leading to $\mathfrak{F}_{n+1} \geq \mathfrak{F}_n \geq \mathfrak{F}_0$, where $n=0$ represents no error reduction. However, in practice, larger circuits introduce more noisy gates, limiting the \textit{theoretical} benefits of error reduction via coherent wavefunction interference.
The robustness of the scheme against noise arising from the preparation of the auxiliary qubit, and from cross-talk between different qubits, is discussed in Section~\ref{sec:noiseenc}.

\subsection{Optimization scheme}

As discussed in Section \ref{subsec:gates}, circuits with $n=1, 2, 3$ ancillas require 15, 72, and 312 parameters, respectively. While optimization with one ancilla is straightforward, the exponential increase in parameters with additional ancillas requires substantial computational resources.
To address this, we utilized the ReCaS-Bari Data Center~\cite{Recas}, a high-performance computing (HPC) cluster with 8000 cores capable of running parallel jobs. Circuit optimization was performed using \texttt{PennyLane}, a Python library tailored for machine learning techniques suited to large parameter spaces\footnote{Code at \href{https://github.com/giovanniscala/ErrorMitigation}{https://github.com/giovanniscala/ErrorMitigation}.}.

For optimization, the best results are achieved using either gradient descent and gradient-free optimizers, with convergence monitored to determine the number of iterations. Multiple runs with different random seeds ensured stability and consistency. This approach enhances the performance and reliability of quantum circuits.

\subsection{Entanglement fidelity}

Here we specialize the scheme in Fig.~\ref{circuit} to the goal of optimizing the entanglement fidelity (see Fig.~\ref{circuit_ef}). 
We denote the two qubits of the initial states with $(R)$ the reference qubit that is not transmitted through the channel and with $(S)$ the signal qubit that is affected by the noise, such that
\begin{align}\label{initialEF}
    |\Phi^+\rangle =
    \frac{1}{\sqrt{2}}
    \left(
    |0\rangle_R |0\rangle_S
    +
    |1\rangle_R |1\rangle_S
    \right) \, .
\end{align}

At the output, the state $\rho_U^{\mathrm{out}}$ is conditioned on measuring the $n$ ancillary qubits in the computational state $|0\rangle^{\otimes n}$. In fact, $\rho_U^{\mathrm{out}} = \rho^\text{out}_{|0\rangle^{\otimes n}}/P_n$ represents the normalized density matrix obtained after post-selection, where $P_n$ is the corresponding post-selection probability,
\begin{align}\label{eq:Pn}
    P_n = \mathrm{Tr} \, \rho^\text{out}_{|0\rangle^{\otimes n}} \, .
\end{align}
The success probability $P_n$ is independent of the cost function. However, in favour of readability, we always plot $P_n$ alongside the cost function.
The entanglement fidelity then reads
\begin{align}
    \mathcal{F}_n = \langle \Phi^+| \rho_U^{\mathrm{out}} | \Phi^+ \rangle
    \, .
\end{align}

\begin{figure}[t]
\resizebox{.95\linewidth}{!}{
\begin{quantikz}[row sep={0.8cm,between origins}, column sep=0.4cm]
\lstick[2]{$\ket{\Phi^+}$} & \qw & \qw& \qw & \qw & \qw & \qw \rstick[2]{$\rho_U^{\mathrm{out}}$} \\
\lstick{}&\qw &\qw  & \gate[wires=2,style={fill=green!20}]{\, U \, } & \gate[style={fill=red!20}]{ \hspace{0.2cm} \mathcal{E} \hspace{0.2cm} } & \gate[wires=2,style={fill=green!20}]{U^\dagger} & \qw \\
\lstick[1]{$\ket{0}^{\otimes n}$}&\qw & \qwbundle{n} & \qw & \gate[style={fill=red!20}]{\mathcal{E}^{\otimes n}} & \qw & \meter{} \rstick{$ \sigma_z^{\otimes n}$}
\end{quantikz}
}
\caption{
Error filtration scheme: The reference and signal qubits are prepared in the maximally-entangled two-qubit state $|\Phi^+\rangle$. 
The signal qubit interacts with $n$ ancillary qubits via the encoding unitary $U$, then passes through a noisy channel $\mathcal{E}$ with (iid) errors.
The inverse unitary followed by $\sigma_z$ measurements on each ancilla decode the post-selected state $\rho_U^{\mathrm{out}}$.
}
\label{circuit_ef}
\end{figure}

\begin{figure}[t]
    \centering
    \includegraphics[width=1\linewidth]{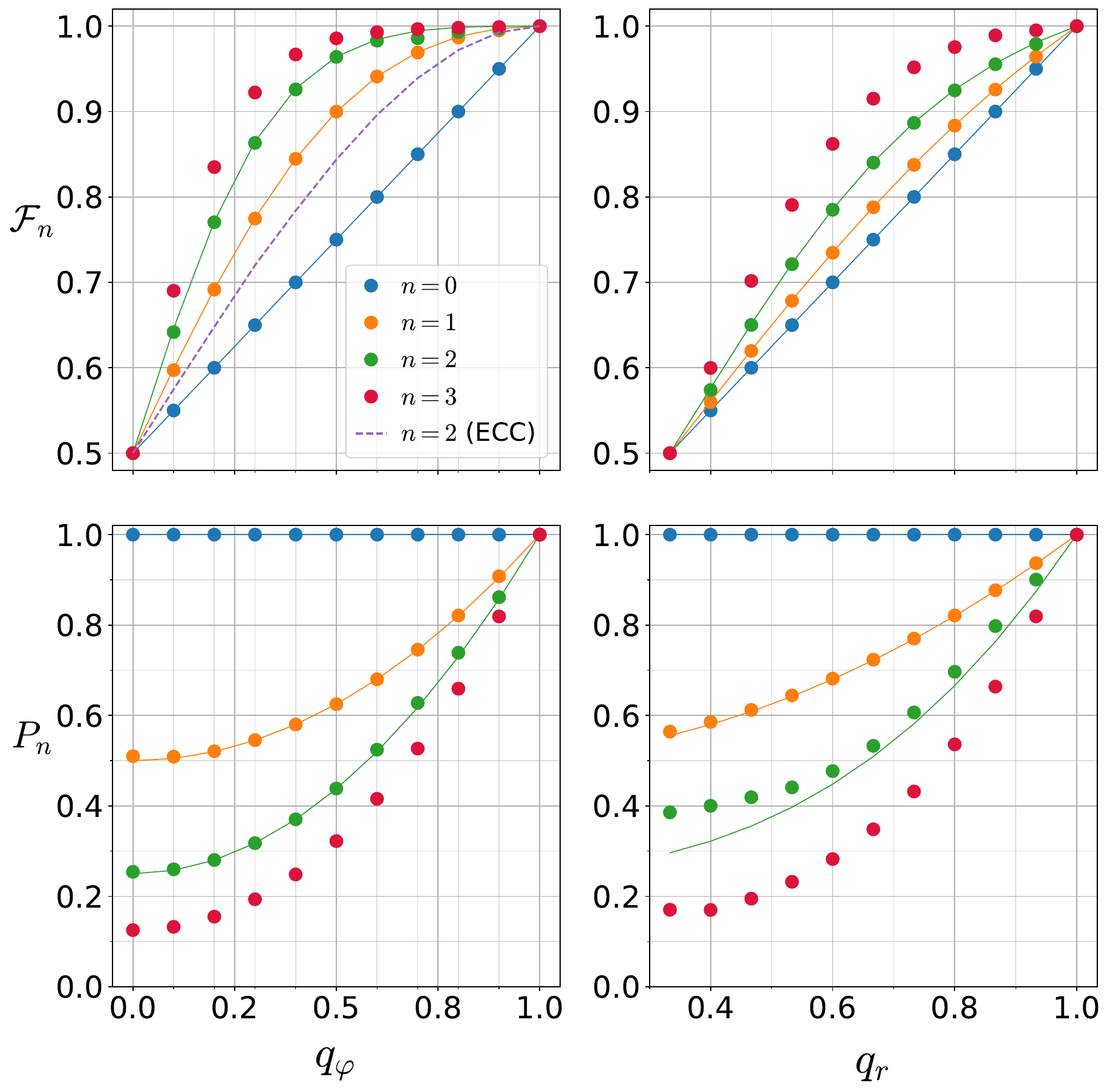}
    \caption{
Entanglement fidelity (upper panel) and success probability (lower panel) following error filtration, as functions of the noise parameters $q_\varphi \in [0, 1]$ for the \textit{dephasing channel} [Eq.~(\ref{eq:state_dephase_noise})] and $q_r \in [1/3, 1]$ for the \textit{depolarizing channel} [Eq.~(\ref{state_depolarizing_noise})] with $n = 0, 1, 2, 3$ ancillas. 
Numerical results are shown as dots.
The solid lines show the analytical results obtained using the code words in Eq.~(\ref{encoding0})-(\ref{encoding1}) and (\ref{2encoding0})-(\ref{2encoding1}).
For comparison, the dashed line in the top-left plot shows the optimized entanglement fidelity achievable by deterministic error correcting code (ECC) for $n = 2$ (see Section~\ref{sec:ecc} and Fig.~\ref{circuit_efV}).
For $n=1$, deterministic error correction is not capable of increasing the fidelity above the $n=0$ value. }%
    \label{fig:fidelity}
\end{figure}

The results of the numerical optimization~\footnote{Here we optimize over the unitary operation, parameterized using QSD. We believe that this is a natural choice since we are optimizing the fidelity of the conditional state 
$\rho_U^{\mathrm{out}} = \rho^\text{out}_{|0\rangle^{\otimes n}}/P_n$, which has a non-linear dependence on $U$ seen as a completely positive map. A more common approach to optimize the fidelity, within quantum error correction, exploits semidefinite programming~\cite{fletcher2008structured}, where the state of interest is not conditional (since error correction is deterministic) and therefore the functional is no longer non-linear. In principle this approach may also be applied to error filtration by relaxing the unitary to a completely positive map, and by accounting for the probability of success $P_n$ through a Lagrange multiplier.} are shown by the circle data points in Fig.~\ref{fig:fidelity} and compared with the fidelity attainable without error filtration, which is $\mathcal{F}_0 = (1 + q_\varphi)/2$ for the dephasing channel, and
$\mathcal{F}_0 = (1 + 3 q_r)/4$ 
for the depolarizing channel
(without error filtration this is achieved deterministically, i.e.~with $P_0=1$).
Figure~\ref{fig:fidelity} shows the trade-off arising when introducing ancillary qubits:
with more ancillas the fidelity is enhanced but at the cost of reducing the success probability. 
We also note that the scheme works better for dephasing rather than depolarizing noise. This is in line with the observations of Refs.~\cite{PhysRevLett.131.230601,PhysRevA.108.062604} and corresponds to the fact that the depolarizing channel has higher Kraus rank (i.e., the number of Kraus operators) than the dephasing channel.

\begin{figure}[htbp]
\resizebox{.95\linewidth}{!}{
\begin{quantikz}[row sep={0.8cm,between origins}, column sep=0.4cm]
\lstick[2]{$\ket{\Phi^+}$} & \qw & \qw & \qw & \qw & \qw & \qw \rstick[2]{$\rho_{U,V}^{\mathrm{out}}$} \\
\lstick{} & \qw & \qw & \gate[wires=2,style={fill=green!20}]{\, U \, } & \gate[style={fill=red!20}]{ \hspace{0.2cm} \mathcal{E} \hspace{0.2cm} } & \gate[wires=2,style={fill=blue!20}]{V\hspace{0.1cm}} & \qw \\
\lstick[1]{$\ket{0}^{\otimes n}$} & \qw & \qwbundle{n} & \qw & \gate[style={fill=red!20}]{\mathcal{E}^{\otimes n}} & \qw & \qw 
\end{quantikz}
}
\caption{A general error correction scheme is obtained as modification of Fig.~\ref{circuit}. To model error correction, two different unitaries, $U$ and $V$, are employed for encoding and decoding. Furthermore, the ancillas are discarded by tracing them out, so the whole procedure is deterministic.}
\label{circuit_efV}
\end{figure}

\subsection{Relation with error correction}\label{sec:ecc}

To shed some light into the relation between error filtration and error correction, here we compare the two approaches when limited resources are available. 
In this setup, we expect error correction to give only approximate results, especially when the noise is relatively strong.
To model error correction, we modify the circuit of Fig.~\ref{circuit} by letting the decoder be described by a unitary $V\in\mathrm{SU}(2^{n+1})$ independent of the encoding unitary $U$.
Also, after the noise and the decoding step, we \emph{trace out} the ancillary qubits instead of measuring them.
This deterministic set up, shown in Fig.~\ref{circuit_efV}, does in fact model a generic error correction scheme, as syndrome detection and conditional correcting operations can be coherently implemented within the decoding unitary $V$.
As a figure of merit we evaluate the optimized entanglement fidelity, where for the scheme in Fig.~\ref{circuit_efV} both $U$ and $V$ are independently optimized.

As an example, we compare error filtration and correction for the case of one ancillary qubit.  
In Fig.~\ref{fig:fidelity} the deterministic optimized  entanglement fidelity is shown for $n=2$, see the data labeled as ECC for error correcting code.
The plot shows that the fidelity lies below the corresponding $n=2$ fidelity obtained within error filtration.
For $n=1$, with deterministic error correction one cannot achieve entanglement fidelity above the $n=0$ value.

In conclusion, at the cost of being probabilistic, error filtration may achieve higher fidelity than deterministic error correction. This pattern may be observed in the case of limited resources as small number of ancillary qubits.


\subsection{Relation with error detection codes} \label{sec:general}

In this Section we provide a more general analysis of error filtration, clarifying the features that characterize the optimal encoding unitary.
At first, we consider a general model of memoryless channel model with Kraus decomposition
\begin{align}
    \mathcal{E}(\rho) = \sum_j K_j \rho K_j^\dag \, .
\end{align}
Using $n$ ancillary systems initialized in the state $|0\rangle$, the encoding unitary maps the input states $|0\rangle  |0\rangle^{\otimes n}$, $|1\rangle  |0\rangle^{\otimes n}$ into
\begin{align}
|\psi_0\rangle & =    U |0\rangle  |0\rangle^{\otimes n} \, , \\
|\psi_1\rangle & =    U |1\rangle  |0\rangle^{\otimes n} \, .
\end{align}
To compute the entanglement fidelity, we need to introduce a reference qubit and consider the state
\begin{align}
|\Psi\rangle = \frac{1}{\sqrt{2}}
\left(
|0\rangle |\psi_0\rangle +
|1\rangle |\psi_1\rangle
\right) \, .
\end{align}

For a given realization of the noise, such a state transforms into
\begin{align}\label{onenoise}
\ket{\Psi_{\{j\}}} =
\frac{1}{\sqrt{2}}
\left(
|0\rangle K_{j_1} \cdots  K_{j_n} |\psi_0\rangle +
|1\rangle 
K_{j_1} \cdots  K_{j_n} |\psi_1\rangle
\right) \, .
\end{align}
Finally, averaging over all noise realizations we obtain
\begin{align}
P_n \mathcal{F}_n = \frac{1}{4}
\sum_{j_1,\dots, j_n}
\left|
\sum_a
\langle \psi_a| K_{j_1} \cdots K_{j_n} |\psi_a\rangle 
\right|^2 \, .
\end{align}

This expression is the product of the entanglement fidelity and the probability of successful post-selection.
The latter is obtained from Eq.~(\ref{onenoise}) by projecting onto the subspace spanned by $|\psi_0\rangle$ and $|\psi_1\rangle$:
\begin{align}
    P_n & =
    \sum_{\{j\}}\mathrm{Tr}[(1_{\mathrm{R}}\otimes \Pi)
    \ket{\Psi_{\{j\}}}
    \bra{\Psi_{\{j\}}}
    (1_{\mathrm{R}}\otimes \Pi)]
    \nonumber\\
    &=\frac{1}{2} 
    \sum_{j_1, \dots , j_n}
    \sum_{a,b}
    \left|
    \langle \psi_a | K_{j_1} \cdots  K_{j_n} |\psi_b\rangle
    \right|^2
    \, .
\end{align}
where $\Pi=\sum_a \ket{\psi_a}\bra{\psi_a}$ is the projector onto the code.
The goal of our optimization routine is to maximize the fidelity
\begin{align}
\mathcal{F}_n = 
\frac{1}{2}
\frac{
\sum_{j_1,\dots, j_n}
\left|
\sum_a
\langle \psi_a| K_{j_1} \cdots K_{j_n} |\psi_a\rangle 
\right|^2 }
{
    \sum_{j_1, \dots , j_n}
    \sum_{a,b}
    \left|
    \langle \psi_a | K_{j_1} \cdots  K_{j_n} |\psi_b\rangle
    \right|^2
}
\, .
\end{align}
Here the numerator represents the ``coherent" part of the evolution inside the code space, and the denominator the ``total action" of the noisy sequence, which includes undesired transitions between codewords. 
First, we may attempt to minimize the denominator.
We have 
\begin{align}
    \sum_{a,b}
    \left|
    \langle \psi_a | K_{j_1} \cdots  K_{j_n} |\psi_b\rangle
    \right|^2
\geq 
    \sum_{a}
    \left|
    \langle \psi_a | K_{j_1} \cdots  K_{j_n} |\psi_a\rangle
    \right|^2
\, .
\end{align}
The lower bound is saturated if the space spanned by $|\psi_a\rangle$, $|\psi_b\rangle$
is an error detection code for the errors $K_{j_1} \cdots  K_{j_n}$, i.e.~it satisfies the error detection condition~\cite{KnillLaflammeViola2000}
\begin{align}\label{EDcond0}
    \Pi K_{j_1} \cdots  K_{j_n} \Pi = c_{j_1 \cdots j_n} \Pi 
    \, .
\end{align}
This means that the action of $K_{j_1},\dots K_{j_n}$ on the code subspace is \textit{proportional to the identity}.
The expression for the fidelity then reads:
\begin{align}
\mathcal{F}_n = 
\frac{
\sum_{j_1,\dots, j_n}
\left|
    c_{j_1 \cdots j_n}
\right|^2 }
{
 \sum_{j_1, \dots , j_n}
    \left|
    c_{j_1 \cdots j_n}
    \right|^2
}
= 1
\, .
\end{align}
Here we have unit fidelity as the matrix elements between distinct code words vanish, yielding no leakage or mixing between codewords.

We conclude that a code that detects all error yields a perfect error filtration scheme. 
In practice, the condition (\ref{EDcond0}) may be satisfied only by a subset of errors.
We thus introduce a milder condition:
\begin{align}\label{milder}
    \langle \psi_a| K_{j_1} \cdots  K_{j_n} |\psi_b \rangle  = 
    \delta_{ab} \,
    c_{a j_1 \cdots j_n} 
    \, ,
\end{align}
meaning that the Kraus operators can act differently on different codewords, but they still do not mix codewords.
Under this condition, the fidelity reads
\begin{align}
\mathcal{F}_n = 
\frac{1}{2}
\frac{
\sum_{j_1,\dots, j_n}
\left|
\sum_a
c_{a j_1 \cdots j_n} 
\right|^2 }
{
    \sum_{j_1, \dots , j_n}
    \sum_{a}
    \left|
    c_{a j_1 \cdots j_n}
    \right|^2
}
\, .
\end{align}

In the case of one ancillary qubit, the code words
\begin{align}
|\psi_0\rangle & = \frac{|00\rangle + |1 1\rangle}{\sqrt{2}} \, , 
\label{encoding0} \\ 
|\psi_1\rangle & =
\frac{|10\rangle - |01\rangle}{\sqrt{2}} \, , 
\label{encoding1}
\end{align}
satisfy condition (\ref{milder}) for all phase-flip errors (on one and two qubits) and for all bit-flip errors (on one and two qubits), but not for combinations of phase-flip and bit-flip errors.
When applied to dephasing channel with parameter $q_\phi$, 
the 
entanglement fidelity 
and the
post-selection probability read
\begin{equation}
    \mathcal{F}_1 (q_\varphi) =\frac{1}{2} + \frac{q_\varphi}{1 + q_\varphi^2}
    \, , 
    \, \, \,
    P_1(q_\varphi) = \frac{1+q_\varphi^2}{2}  
 \, .
     \label{anc1deph}
\end{equation}
When applied to a depolarizing channel of parameter $q_r$ we obtain
\begin{equation}
    \mathcal{F}_1 (q_r) = 
    \frac{1+2q_r+5q_r^2}{4(1+q_r^2)} 
    \, , 
    \, \, \,
    P_1(q_r) = 
    \frac{1+q_r^2}{2} 
     \, .
     \label{anc1depo}
\end{equation}
The above expression for both noise models match the results of the numerical optimizations (see Fig.~\ref{fig:fidelity}).

For two ancillas, similar properties hold for the code words
\begin{align}
|\psi_0\rangle & = 
\frac{|000\rangle + |011\rangle + |101\rangle + |110\rangle  }{2} \, , 
\label{2encoding0} \\
|\psi_1\rangle & =
\frac{ |001\rangle - |010\rangle + |100\rangle - |111\rangle }{2} \, ,
\label{2encoding1}
\end{align}
which satisfy condition (\ref{milder}) for all phase-flip errors (acting on one up to three qubits) and for all bit-flip errors (acting on one up to three qubits), but not for their combinations.
Using this code we obtain, respectively for dephasing and depolarizing noise,
\begin{equation}
\mathcal{F}_2(q_\varphi) = \frac{(q_\varphi +1)^3}{6 q_\varphi^2+2}
    \, , 
    \, \, \,
P_2(q_\varphi) = \frac{1+ 3 q_\varphi^2}{4} \, ,
     \label{anc2deph}
\end{equation}
and
\begin{equation}
 \mathcal{F}_2(q_r) = \frac{1+7q_r^2}{4(1-q_r+2q_r^2)}
    \, , 
    \, \, \,
   P_2(q_r) = \frac{1+q_r^2+2q_r^3}{4} \, .
     \label{anc2depo}
\end{equation}
Also for this code, the entanglement fidelity matches the 
results of the numerical optimization (see Fig.~\ref{fig:fidelity}), hence indicating the relevance of the weak error detection condition in Eq.~(\ref{milder}).

\subsection{Noisy implementation of error filtration}\label{sec:noiseenc}

In this Section, we look at what happens when the encoding procedure is imperfect. 
In practical experiments, the introduction of ancillary qubits and the entangling unitary $U$ inevitably adds noise into the system.
One would expect that the impact of this \textit{filtration-induced} noise may become more noticeable as we increase the number of ancillary qubits. 
However, here we show that there exist noise models for which this detrimental effect remains under control, and in fact, we can still obtain some benefits from increasing the number of ancillas. 

We focus on two main sources of error: (i) ancillary qubits that may themselves be noisy, and (ii) cross-talk during encoding ($U$) and decoding ($U^\dag$).
The former is modeled by adding an extra depolarizing channel to each ancillary qubit, the latter by introducing stochastic swap of signal and ancillas each time a multi-qubit unitary is applied.
Our goal is to see how these imperfections impact the entanglement fidelity.

\subsubsection{Noisy ancilla preparation}

In this first scenario, we introduce a local depolarizing channel acting on each ancillary qubit, just before it enters the filtration circuit. 
The depolarizing channel is denoted $\mathcal{D}$ and is characterized by a noise parameter $q_a$ (see definition in Section~\ref{subsec:noise}).
This models preparation noise of the initial ancillary qubits state.
Even though we aim to protect the signal, errors on the ancillary qubits can still reduce the final fidelity. 
The scheme is depicted in Fig.~\ref{circuit_efn}.
As more ancillae are added, the overall noise can grow, so it is important to see how well our method holds up in this situation.

To address this question, we perform the optimization of the fidelity in the presence of preparation noise.
In the top panel of Fig.~\ref{fig:f_noisy} we plot the computed optimal fidelity against the depolarizing parameter $q_a$ applied to each ancilla qubit. 
When $q_a$ approaches $1$, the fidelity improves and moves closer to its ideal value.
We observe that despite preparation noise, having more ancillary qubits still provides stronger suppression of errors.

\begin{figure}[t]
\resizebox{.95\linewidth}{!}{
\begin{quantikz}[row sep={0.8cm,between origins}, column sep=0.4cm]
\lstick[2]{$\ket{\Phi^+}$}&\qw & \qw & \qw & \qw & \qw & \qw & \qw & \qw \rstick[2]{$\rho_U^{\mathrm{out}}$} \\
 \lstick{} &\qw &\qw  & \qw & \gate[wires=2,style={fill=green!20}]{\, U \, } & \gate[style={fill=red!20}]{ \hspace{0.2cm} \mathcal{E} \hspace{0.2cm}} & \gate[wires=2,style={fill=green!20}]{U^\dagger} & \qw & \qw \\
 \lstick[1]{$\ket{0}^{\otimes n}$}&\qw & \qwbundle{n} & \gate[style={fill=black!20}]{\mathcal{D}^{\otimes n}}{ \hspace{0.2cm}} & \qw & \gate[style={fill=red!20}]{\mathcal{E}^{\otimes n}} & \qw & \qw & \meter{} \rstick{$ \sigma_z^{\otimes n}$}
\end{quantikz}
}
\caption {Error filtration scheme with depolarizing noise in the preparation of the ancillary qubits: the signal qubit interacts with $n$ noisy ancillary qubits before passing through the encoding unitary $U$, and then goes through the main noisy channel $\mathcal{E}$.}
\label{circuit_efn}
\end{figure}

\subsubsection{Cross-talk}

In the second scenario, we model unwanted cross-talk during the application of the encoding ($U$) and decoding ($U^\dag$) multi-qubit unitary. 
The scheme is shown in Fig.~\ref{circuit_sw}.
We model cross-talk during the encoding process by introducing random SWAP operations between the signal qubit and ancillae. 
We model this process as follows. 
In the case of $n=1$ ancilla, with probability $s$ no swap occurs and the encoding remains intact, and with probability $1 - s$ the signal qubit is swapped with the ancilla. 
In the case of $n=2$ ancillae, the signal qubit can swap with the either the first or second ancilla, each occurring with probability $\frac{1 - s}{2}$. 
Likewise, for $n=3$ ancillas, the individual swap probability is $\frac{1 - s}{3}$.

\begin{figure}[tbh]
    
    \includegraphics[width=0.9\linewidth]{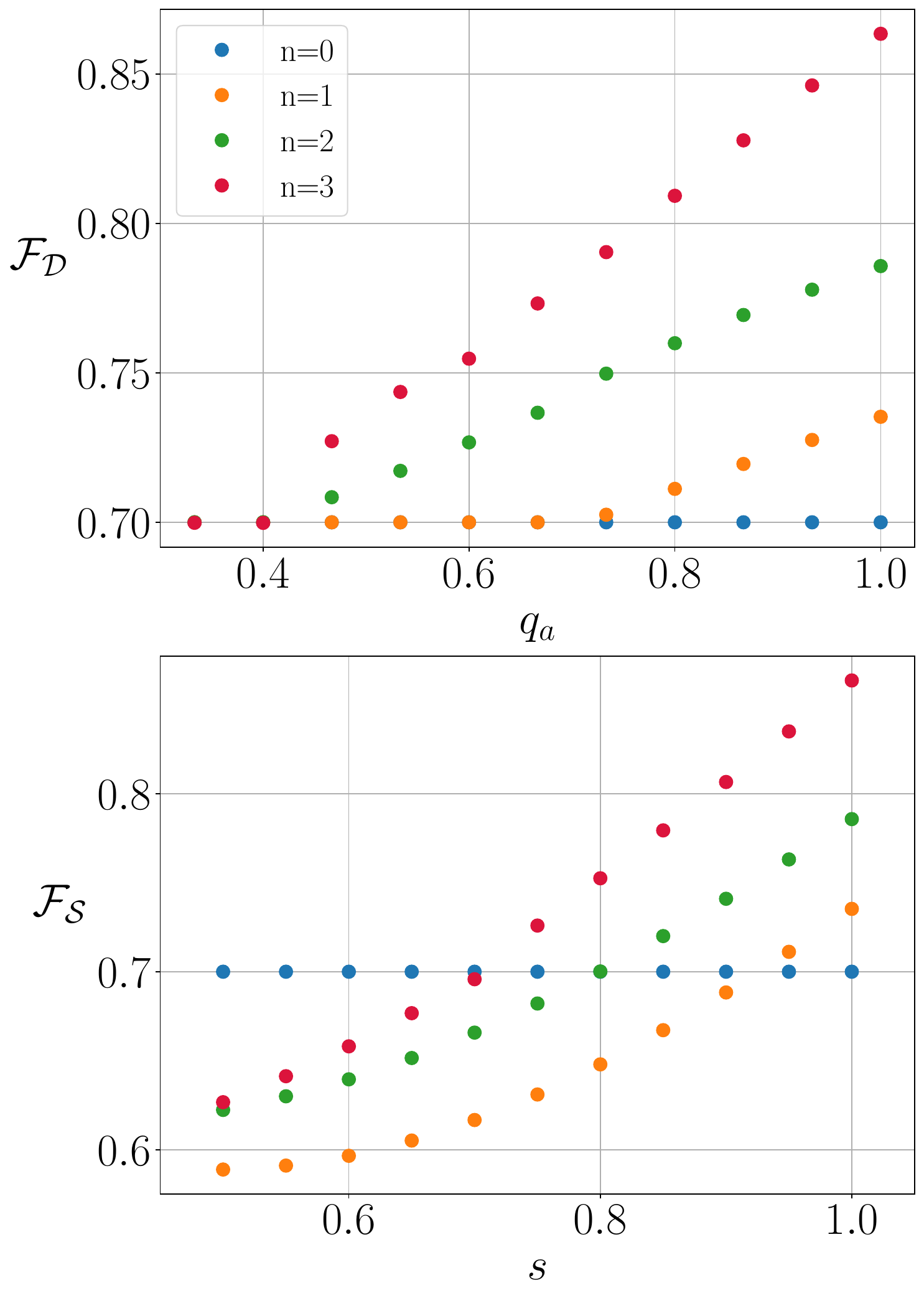}
    \caption{
    Numerical results for the two error models described in Section~\ref{sec:noiseenc}. The top panel shows how the optimized entanglement fidelity
    $\mathcal{F}_\mathcal{D}$ depends on the preparation noise depolarizing parameter $q_a \in \left[\tfrac{1}{3}, 1\right]$, with the main channel noise fixed at $q_r = 0.7$.
    The bottom panel shows how the entanglement fidelity $\mathcal{F}_\mathcal{S}$ changes as functions of the swap probability $s \in [0.5, 1]$, again for a fixed channel noise strength of $q_r=0.7$. 
    Each curve corresponds to a different number of ancilla qubits ($n=0,1,2,3$).}
    \label{fig:f_noisy}
\end{figure}

We compute the optimized fidelity in the presence of this cross-talk model. The results are shown in the bottom panel of Fig.~\ref{fig:f_noisy} by varying the SWAP probability $s$, while keeping the main channel noise fixed. 
As $s$ approaches $1$ (ideal, no cross-talk limit), the fidelity gets its optimal value, and using more ancilla qubits further boosts performance. 
However, the filtration scheme shows some robustness to this noise model as increasing the 
number of ancillas 
systematically counteracts the effects of cross-talk, showing that our scheme remains robust despite noisy pre- and post-processing steps.

\begin{figure}[t]
\resizebox{.95\linewidth}{!}{
\begin{quantikz}[row sep={0.8cm,between origins}, column sep=0.4cm]
\lstick[2]{$\ket{\Phi^+}$} & \qw & \qw & \qw & \qw & \qw & \qw & \qw & \qw \rstick[2]{$\rho_U^{\mathrm{out}}$} \\
 \lstick{} & \qw & \gate[wires=2,style={fill=black!20}]{\, S \, } & \gate[wires=2,style={fill=green!20}]{\, U \, } & \gate[style={fill=red!20}]{ \hspace{0.2cm} \mathcal{E} \hspace{0.2cm}} & \gate[wires=2,style={fill=green!20}]{U^\dagger} & \gate[wires=2,style={fill=black!20}]{\, S \, } & \qw & \qw \\
 \lstick[1]{$\ket{0}^{\otimes n}$} & \qwbundle{n} & \qw & \qw & \gate[style={fill=red!20}]{\mathcal{E}^{\otimes n}} & \qw & \qw & \qw & \meter{} \rstick{$ \sigma_z^{\otimes n}$}
\end{quantikz}
}
\caption{Error filtration scheme with cross-talk between the signal and ancillary qubits:
a random SWAP is applied between the signal qubit and ancillae at the encoding stage before $U$. 
The state then goes through the main channel $\mathcal{E}$ and another random SWAP applies again after it is decoded by $U^\dagger$.}
\label{circuit_sw}
\end{figure}

\begin{figure}[t]
    \centering
    \includegraphics[width=0.9\linewidth]{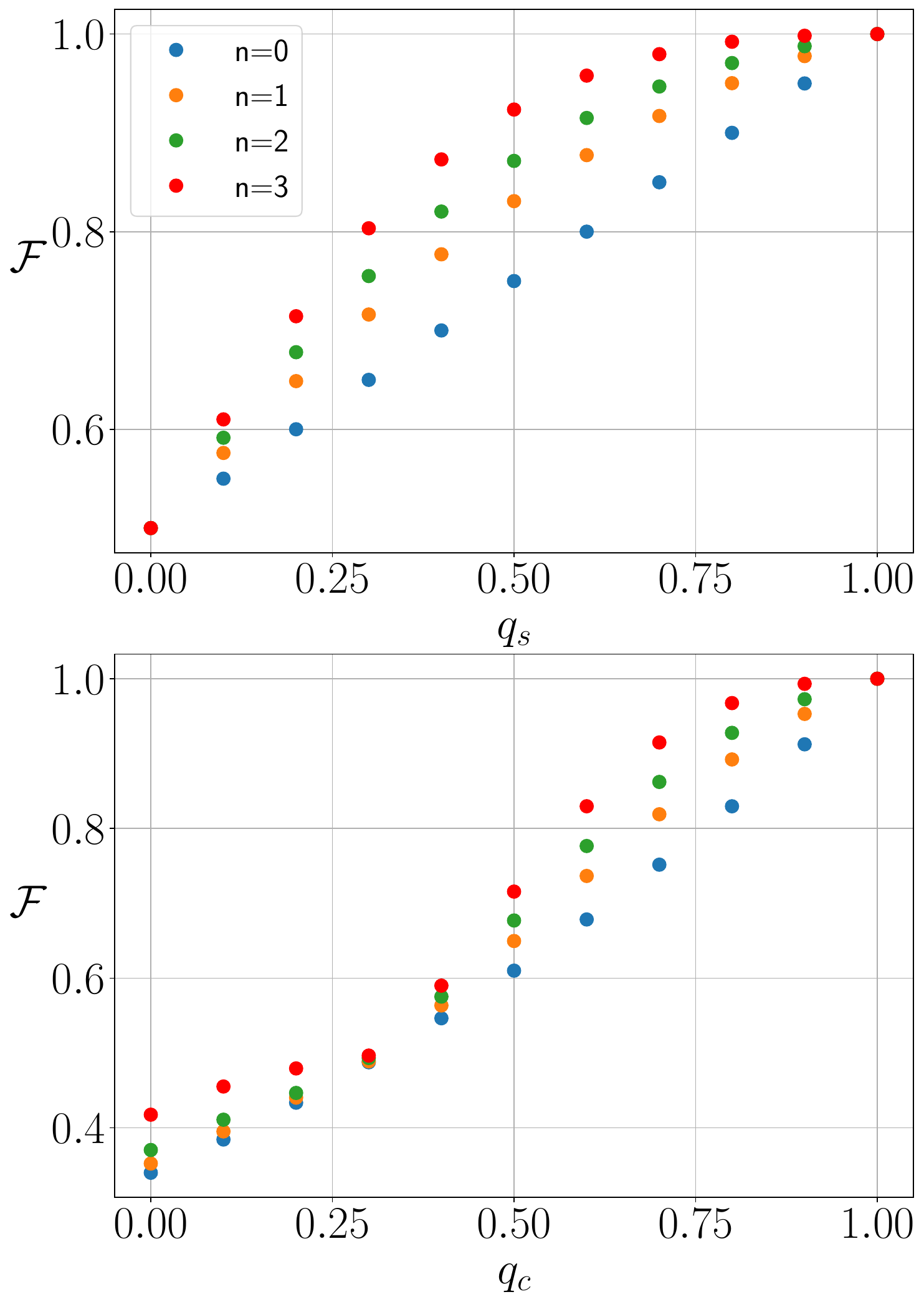}
    \caption{Entanglement fidelity  under composite noisy model of Eq.~\eqref{eq:cnoisy} (lower panel) for $n=0,1,2,3$ ancilla for $q_\varphi=0.3$, $q_r=0.6$, $q_{\mathrm{SWAP}}=0.1$; the stochastic Pauli noise model (upper panel) for $q_x=0.5$, $q_y=0.3$, $q_z=0.2$.}
    \label{fig:hybrid_fidelity}
\end{figure}


Figure~\ref{fig:hybrid_fidelity} shows the protocol’s effectiveness in filtering more complex noise models.
We probabilistically combine different models considered above, namely dephasing, depolarizing and cross--talk:
\begin{equation}\label{eq:cnoisy}
    \mathcal{E}_c=q_c\mathrm{id} +(1-q_c)(q_\varphi \mathcal{E}_\varphi+q_r \mathcal{E}_r+q_{\mathrm{SWAP}} \mathcal{E}_\mathrm{SWAP}). 
\end{equation}
In particular, upper panel of Fig.~\ref{fig:hybrid_fidelity} specializes the depolarizing effect with non-uniform shrinking factors $q_x,q_y,q_z$ instead of $q_r$ (or ($p_r$) of Eq.~\eqref{eq:depolarizing_noise}, a.k.a. stochastic Pauli noise model,
\begin{equation}\label{eq:sPnoisy}
    \mathcal{E}_s(\,\cdot)=q_s\mathrm{id}(\,\cdot) +(1-q_s)\sum_{k=x,y,z}q_k \sigma_k(\,\cdot) \sigma_k.
\end{equation}

\subsection{Violation of Bell inequalities}

The main challenge in implementing device-independent quantum key distribution~\cite{Nadlinger2022,Liu2022,Zhang2022a}
is reducing noise to violate suitable Bell inequalities~\cite{Mandarino2023}. Therefore, we analyse whether our method is useful for this application.
In a simple Bell scenario, a bipartite state $\rho_{AB}$ of two qubits is shared between two distant labs, Alice (A) and Bob (B). 
In each lab, one of two possible observables is measured, represented by matrices $M_{i,j}$, where 
\begin{equation}
    M_{i,j}=\begin{pmatrix}
\cos\theta_{ij} & e^{-i \phi_{ij}} \sin\theta_{ij} \\
 e^{i \phi_{ij}} \sin\theta_{ij} & - \cos\theta_{ij}
\end{pmatrix}
.
\end{equation}
Here $i=0,1$ labels the sub-systems ($i=0$ for Alice and $i=1$ for Bob), and $j=0,1$ labels two possible measurement settings.
The measurements are dichotomic, i.e., each has two possible outcomes $a,b\in\{-1,+1\}$. 
After repeated measurements, Alice and Bob can estimate the CHSH functional~\cite{CHSH69andSangouard}
\begin{equation}\label{eq:Bell_functional}
\mathcal{B} = 
\left|
\sum_{x,y=0}^1(-1)^{xy}
\,
\mathrm{Tr}\left(M_{0,x}\otimes M_{1,y} \, \rho_{AB}
\right)
\right| 
\, .
\end{equation}
According to Bell's theorem, a local realistic theory satisfies $\mathcal{B} \leq 2$. However, if $\rho_{AB}$ is maximally entangled~\cite{cirel1980quantum}, this bound increases to $\mathcal{B} = 2\sqrt{2}$, allowing for secure communication provided $2 < \mathcal{B} \leq 2\sqrt{2}$. 
However, when $\rho_{AB}$ is distributed between Alice and Bob, as in Fig.~\ref{circuit_ef2}, noise in the communication line reduces the value of $\mathcal{B}$~\cite{abellan2018,Gigena2022,handsteiner2017,rauch2018,scheidl2010,aktas2015}. To counteract this, we apply the error filtration technique to maintain secure communication by maximizing $\mathcal{B}$.

\begin{figure}[t]
\resizebox{.9\linewidth}{!}{
\begin{quantikz}[row sep={0.8cm,between origins}, column sep=0.4cm]
\lstick[2]{$\ket{\Phi^+}$} & \qw & \qw & \qw & \qw & \gate{ \hspace{-0.07cm} M_{0,x} \hspace{-0.07cm} } \\
 \lstick{} & \qw & \gate[wires=2,style={fill=green!20}]{\, U \, } & \gate[style={fill=red!20}]{ \hspace{0.2cm} \mathcal{E} \hspace{0.2cm} } & \gate[wires=2,style={fill=green!20}]{U^\dagger} & \gate{ \hspace{-0.07cm} M_{1,y} \hspace{-0.07cm} } \\
 \lstick[1]{$\ket{0}^{\otimes n}$} & \qwbundle{n} & \qw & \gate[style={fill=red!20}]{\mathcal{E}^{\otimes n}} & \qw & \meter{}
\end{quantikz}
}
\caption{
Error filtration scheme designed to maximize CHSH inequality violation. A maximally-entangled two-qubit state $|\Phi^+\rangle$ is prepared in Alice's lab, where she stores the first qubit without noise. The second qubit interacts with $n$ ancillary qubits via the encoding unitary $U$ and travels to Bob's lab through a noisy channel $\mathcal{E}$. Bob applies the inverse unitary and measures the ancillary qubits in the computational basis. Alice and Bob perform the Bell test using local operators $M_{0,x}$ and $M_{0,y}$, conditioned on the ancillary qubits being in the state $|0\rangle^{\otimes n}$.
}
\label{circuit_ef2}
\end{figure}

\begin{figure}[t]
    \centering
    \includegraphics[width=0.9\linewidth]{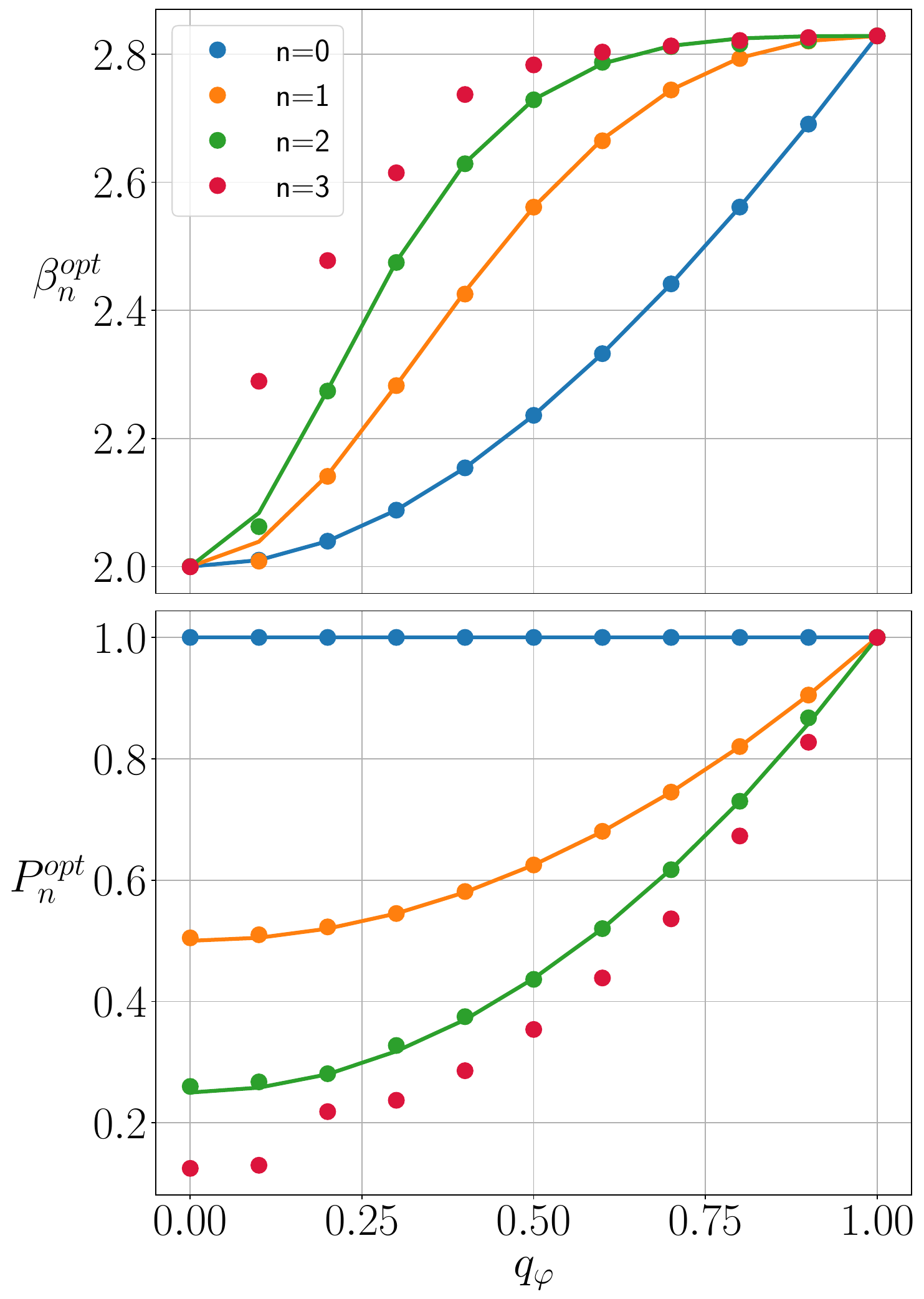}
    \caption{
    Optimal CHSH value (top panel) optimizing also on the measurement settings [see~Eq.~\eqref{eq:beta_opt}] and successful probability $P_n$ (lower panels), plotted vs the noise parameter $q_\varphi \in[0.5,1]$ for \textit{dephasing channel}.
    Dots are for numerical optimization, solid lines for the ansatz unitary, for $n=0,1,2,3$ ancillary qubits.
    }
    \label{fig:CHSH_optimal}
\end{figure}

\begin{figure}[t]
    \centering
    \includegraphics[width=1\linewidth]{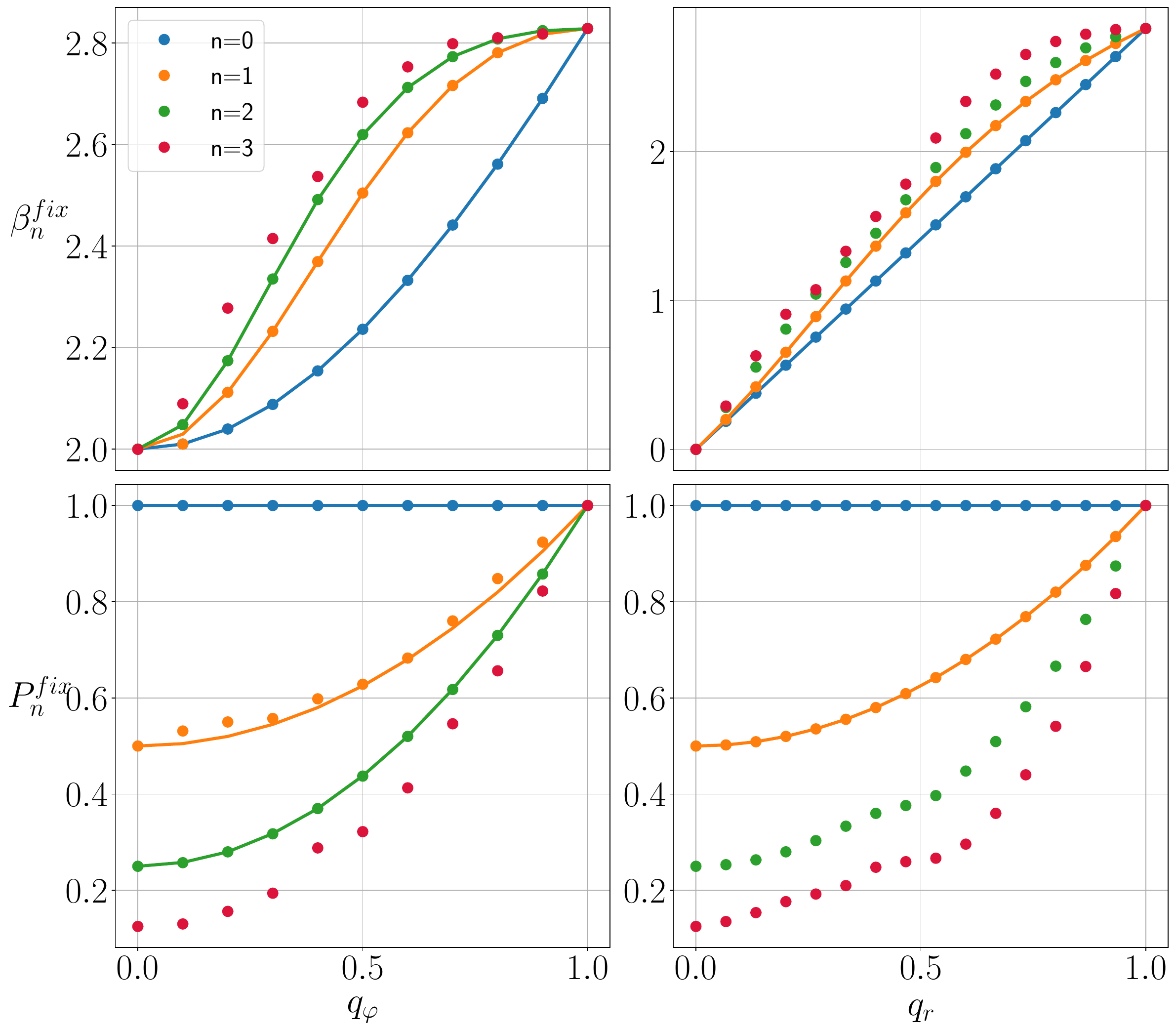}
 \caption{CHSH value (upper panel) with fixed measurement settings and success probability (lower panel) as functions of the noise parameters $q_\varphi \in [0, 1]$ for the \textit{dephasing channel} [Eq.~\eqref{eq:state_dephase_noise}] and $q_r \in [1/3, 1]$ for the \textit{depolarizing channel} [Eq.~\eqref{state_depolarizing_noise}] with $n = 0, 1, 2, 3$ ancillae. Numerical results are shown as dots, and analytical results as solid lines [Eqs.~\eqref{beta2deph}, \eqref{beta1depo}].
 }
 \label{fig:CHSH_fixed}
\end{figure}

From Fig.~\ref{circuit_ef2}, first, Alice locally prepares the entangled state $|\Phi^+\rangle$. Noise (either dephasing or depolarizing) is applied to the qubit transmitted to Bob, with error reduction aided by $n$ ancillary qubits and an encoding/decoding unitary $U$. The ancillas interact with the signal qubit before the noisy channel, and both are subject to independent errors, followed by recombination using $U^\dag$. 
By invoking the fair sampling assumption, the functional in
Eq.~(\ref{eq:Bell_functional}) is computed on the normalized state $\rho_{AB}=\rho^\text{out}_{|0\rangle^{\otimes n}}/P_n$.
This functional depends on parameters defining the unitary $U$ (Section \ref{sec:univ}) and the measurement settings $\theta_{ij}$ and $\phi_{ij}$. We simulate two possible experimental scenarios: finding the optimal unitary for fixed $\theta_{ij}$ and $\phi_{ij}$:
\begin{equation}\label{eq:beta_fix}
\beta^{\mathrm{fix}}_n = \max_{U} \,  \mathcal{B}_n
 \, ,
\end{equation}
and maximizing over both the unitary and measurement angles:
\begin{equation} \label{eq:beta_opt}
\beta^{\mathrm{opt}}_n = \max_{U,\theta_{ij},\phi_{ij}} \, \mathcal{B}_n \, .
\end{equation}
For the estimation of $\beta^{\mathrm{fix}}_n$ we fix the values of $\theta_{ij}$,
$\phi_{ij}$ that maximize the CHSH functional with no error filtration applied (no ancilla). 
Given $\ket{\Phi^+}$, and for the depolarizing channel, these values are well-known:
\begin{align}\label{eq:optimalsettings}
    (\phi_{0,0},\theta_{0,0}) & = (0,0), \quad 
    &(\phi_{0,1},\theta_{0,1}) = \left(0,\frac\pi2\right), \nonumber\\
    (\phi_{1,0},\theta_{1,0}) & = \left(0,\frac\pi4\right), \quad 
    &(\phi_{1,1},\theta_{1,1}) = \left(\pi,\frac\pi4\right) \, .
\end{align}
Similarly, the optimal settings for dephasing noise with parameter $q_\varphi$ are obtained for
\begin{align}\label{eq:optimalsettings2}
    (\phi_{0,0},\theta_{0,0}) & = \left(0,0\right),\,\qquad
    (\phi_{0,1},\theta_{0,1}) = \left(0,\frac\pi2\right), \\
    (\phi_{1,0},\theta_{1,0}) & =  (0,\arctan{q_\varphi}),\,
 (\phi_{1,1},\theta_{1,1})  =  \left( \pi,\arctan{q_\varphi}\right).\nonumber
\end{align}
Using these optimal settings, the maximum value of the Bell functional in the presence of dephasing and depolarizing noise, without our reduction scheme, are well-established~\cite{HORODECKI1995340}. They are given by:
\blk
\begin{equation}
   \beta^\mathrm{fix}_0 (q_\varphi) 
   = 2 \sqrt{ 1 + q_\varphi^2} \, ,\qquad 
\beta^\mathrm{fix}_0 (q_r) = 2 \sqrt{2} \, q_r \, .
\end{equation}
Using the code words in Eq.~(\ref{encoding0})-(\ref{encoding1}) and (\ref{2encoding0})-(\ref{2encoding1}), we obtain the following analytical expressions for the CHSH value after error filtration. 
For dephasing noise ($n=1,2$) 
\begin{equation}
    \beta^\mathrm{fix}_1 (q_\varphi) =\frac{6 q_\varphi^2+2}{\left(q_\varphi^2+1\right)^{3/2}},\quad 
    \beta^\mathrm{fix}_2 (q_\varphi) =\frac{2 (1 + 6 q_\varphi^2 + q_\varphi^4)}{(1 + 3 q_\varphi^2) \sqrt{1 + q_\varphi^2}} \,  ,
    \label{beta2deph}
\end{equation}
and for depolarizing noise ($n=1$)
\begin{align}\label{beta1depo}
 \beta^\mathrm{fix}_1 (q_r) =   2 \sqrt{2} \, q_r \frac{ 1 + q_r }{ 1 + q_r^2 }.
\end{align}
These are the solid lines shown in Fig.~\ref{fig:CHSH_fixed}  that overlap with the numerical optimization, suggesting that the above code words are in some cases also optimal for the CHSH functional. 

Finally, the Bell functional in Eq.~(\ref{eq:beta_opt}) includes also the optimization over measurement settings. In this case, the symmetry of the depolarizing noise implies that the optimal angles remain unchanged.

Instead, for the dephasing noise, optimizing over the angles improves the CHSH value,
$\beta^{\mathrm{opt}}_n>\beta^{\mathrm{fix}}_n$.
This is shown in Fig.~\ref{fig:CHSH_optimal}, along with the matching values obtained 
using Eq.~(\ref{encoding0})-(\ref{encoding1}) and (\ref{2encoding0})-(\ref{2encoding1}).

\subsection{Quantum metrology}

In this Section we apply error filtration to mitigate noise in parameter estimation.
Given a family of states $\rho_\theta$ depending on the continuous parameter $\theta$, the variance $\Delta^2\tilde\theta$ of any unbiased estimator $\tilde\theta$ is bounded as 
$\Delta^2\tilde\theta \ge M^{-1}Q[\rho_\theta]^{-1}$ (quantum Cramér-Rao (QCR) bound) where $M$ is the number of measured copies of $\rho_\theta$, 
\begin{equation}\label{eq:QFI}
Q[\rho_\theta] = \sum_{j,k:\lambda_j+\lambda_k\neq 0}
\frac{2}{\lambda_j+\lambda_k} \, \left| \bra{e_j} \partial \rho_\theta / \partial \theta \ket{ e_k} \right|^2
\end{equation} 
is the quantum Fisher information (QFI)~\cite{paris2009quantum,Sidhu}, 
and
\begin{equation}\label{eq:QFI-spectral}
    \rho_\theta = \sum_j \lambda_j
    \left| e_{j} \right\rangle \left\langle e_{j} \right|
\end{equation}
is the ($\theta$-dependent) spectral decomposition.
The QFI is often employed as a figure of merit in quantum estimation theory. Here we use the QFI as a cost function to be maximized using the technique of error filtration. 

\begin{figure}[t]
\resizebox{.9\linewidth}{!}{
\begin{quantikz}[row sep={0.6cm,between origins}, column sep=0.5cm]
     \lstick[1]{$\ket{\psi^\text{in}_\theta}$} & \qw & \gate[wires=2,style={fill=green!20}]{\, U \, }  & \gate[style={fill=red!20}]{ \hspace{0.2cm} \mathcal{E} \hspace{0.2cm} } & \gate[wires=2,style={fill=green!20}]{ U^\dagger } & \qw  \rstick{$\rho^\text{out}_U$} \\
     \lstick[1]{$\ket{0}^{\otimes n}$} &  \qwbundle{n} & \qw & \gate[style={fill=red!20}]{\mathcal{E}^{\otimes n}}  &\qw& \meter{}
\end{quantikz}
}
\caption{
Error filtration scheme applied to parameter estimation.
The signal qubit is initialized in
$|\psi_\theta^\text{in}\rangle$ $= \cos(\theta/2)|0\rangle$ 
$+ \sin(\theta/2)|1\rangle$ 
(with $\theta$ the parameter to be estimated) and, together with $n$ ancillary qubits in $|0\rangle^{\otimes n}$, is encoded via $U$. The encoded state is then subjected to the local noisy channel $\mathcal{E}$, deocding by $U^\dag$ and post-selection.
The QFI is then evaluated on the state in Eq.~(\ref{tauQFI}).
}
\label{fig:QFI_true}
\end{figure}


\begin{figure}[!htbp]
    \centering
    \includegraphics[width=0.9\linewidth]{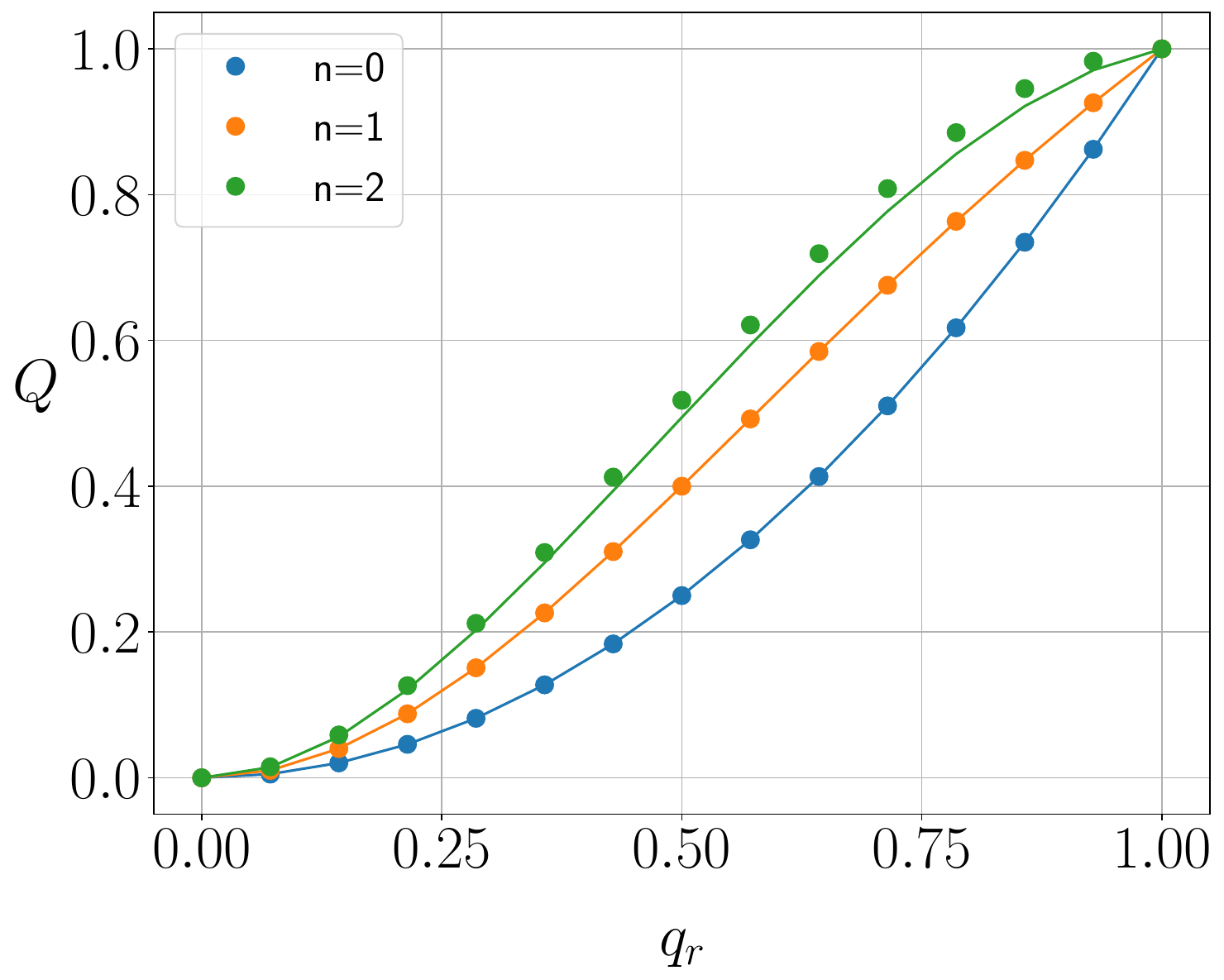}
    \caption{
    The plot shows the optimal value of QFI after error filtration, following the optimization over all encoding unitary with $n=0,1,2$ ancillary qubits.
    The QFI is plotted vs the noise parameter of the depolarizing channel, see  Eq.~\eqref{eq:depolarizing_noise}.
    For $n=0$ (no error filtration), one has $Q_0(q_r) = q_r^2$ (blue solid line). For $n=1$, the orange solid line plots Eq.~(\ref{eq:Q1}), and for $n=2$, the green solid line in Eq.~\eqref{eq:Q2}.
}
    \label{fig:QFI_optimal}
\end{figure}

The error filtration scheme is shown in Fig.~\ref{fig:QFI_true}. 
As a case study, we consider a phase estimation problem, where the parameter 
$\theta$ is initially encoded as
\begin{align} \label{inputangle}
\ket{\psi^\mathrm{in}_\theta} = 
e^{-i\theta \sigma_y/2} |0\rangle =
\cos(\theta/2)|0\rangle + \sin(\theta/2) |1\rangle
\, .
\end{align}
Let 
$\rho_U^{\mathrm{out}} = \rho^\text{out}_{|0\rangle^{\otimes n}}/P_n$ 
be the normalized output state, conditioned on successful post-selection.
Post-selection can be modeled as an erasure channel, which erases the output state with probability $1-P_n$. Therefore, the QFI is computed on the state 
\begin{align}
\tau_\theta & = P_n \rho_U^{\mathrm{out}}
    + (1-P_n) |\omega\rangle \langle \omega| \\
    & = \rho^\text{out}_{|0\rangle^{\otimes n}}
    + (1-P_n) |\omega\rangle \langle \omega| \, ,
    \label{tauQFI}
\end{align}
where $\omega$ is a $\theta$-independent erasure flag state. Note that in general $P_n$ may depend on $\theta$.
Therefore, for quantum metrology applications, the functional to be optimized is
$Q[\tau_\theta]$.

The results of the numerical optimization are shown in Fig.~\ref{fig:QFI_optimal}, for the case of depolarizing noise.
We note that for dephasing noise the error-filtration scheme yields $Q=1$ even without ancillas, as the phase shift $\theta$ around $\sigma_y$ is unaffected by dephasing along $\sigma_z$. 
In fact, if dephasing occurs along $\sigma_y$, a one-qubit rotation mapping $\sigma_y$ to $\sigma_z$ restores $Q=1$.

For $n=1$, we have found an optimal encoding mapping 
$|\psi_\theta^\text{in}\rangle$ into
\begin{align}\label{gfd3h1}
|\Psi_\theta\rangle = \cos{ (\theta/2) }|\psi_0\rangle +
\sin{ (\theta/2) }|\psi_1\rangle \, ,
\end{align}
where
\begin{align}
|\psi_0\rangle & = \frac{|00\rangle + |1 1\rangle}{\sqrt{2}} \, , 
\label{encoding0_QFI} \\ 
|\psi_1\rangle & =
\frac{|10\rangle + |01\rangle}{\sqrt{2}} \, . 
\label{encoding1_QFI}
\end{align}
[Note the difference with Eqs.~(\ref{encoding0})-(\ref{encoding1}); encodings that are optimal for the entanglement fidelity might not be optimal also for the QFI.]
The code (\ref{encoding0_QFI})-(\ref{encoding1_QFI}) is optimal for $\theta=0$, yielding 
\begin{align}\label{eq:Q1}
    Q_1(q_r) = \frac{2 q_r^2}{q_r^2+1} \, ,
\end{align}
which matches the results of the numerical optimization in Fig.~\ref{fig:QFI_optimal}.
For $\theta \neq 0$, the optimality is achieved by rotating the code words by angle $-\theta$. 
This rotation is in fact included in the numerical optimization and operationally corresponds to an optimal adaptive estimation protocol.

For $n=2$, the code words 
\begin{align}
|\psi_0\rangle & = 
\frac{|000\rangle + |011\rangle + |101\rangle + |110\rangle  }{2} \, , 
\\
|\psi_1\rangle & =
\frac{ |001\rangle + |010\rangle + |100\rangle + |111\rangle }{2} 
\end{align}
[again, note the sign difference with Eqs.~(\ref{2encoding0})-(\ref{2encoding1})]
yield, for $\theta=0$,
\begin{align}\label{eq:Q2}
    Q_2(q_r) = q_r^2\frac{23+22 q_r^2-13q_r^4}{8( 3q_r^2 + 1)}  \, .
\end{align}
As also shown in Fig.~\ref{fig:QFI_optimal}, this result
nearly matches the outcomes of the numerical optimizations. 
This leaves open the problem of finding exact optimal codes  and analytical expressions for the optimized QFI with $n=2$.

\begin{figure}[t]
\resizebox{.9\linewidth}{!}{
\begin{quantikz}[row sep={0.6cm,between origins}, column sep=0.5cm]
     \lstick[1]{$\ket{\psi^\text{in}_\theta}$} & \qw & \gate[wires=2,style={fill=green!20}]{\, U \, }  & \gate[style={fill=red!20}]{ \hspace{0.2cm} \mathcal{E} \hspace{0.2cm} } &  \qw \rstick[2]{$\chi_\theta$} \\
     \lstick[1]{$\ket{0}^{\otimes n}$} &  \qwbundle{n} & \qw & \gate[style={fill=red!20}]{\mathcal{E}^{\otimes n}} & \qw
\end{quantikz}
}
\caption{
Error correction scheme for parameter estimation.  
This is analogous to Fig.~\ref{circuit_efV}, but here the decoding unitary $V$ is omitted since the QFI is invariant under unitary transformations.
}
\label{circuit2EC}
\end{figure}

It is interesting to compare to the performance of deterministic error correction in this context.
Other works have addressed error correction within quantum metrology (a summary of some previous works is reported in Table~\ref{tab:qfi}). 
However, in most previous works the parameter to be estimated arises from a noisy dynamics, e.g.~\cite{PhysRevLett.112.150802,zhou2018achieving,gorecki2020optimal}. By contrast, in our model, the parameter $\theta$ arises from a pure, unitary phase shift, and depolarizing noise acts after the phase shift, modeling noisy storage or transmission. 
The main difference is that in our case it makes sense to assume that the same depolarizing noise affects the ancillary qubits, whereas in other works the ancillas are assumed noiseless.

Within our model, the application of quantum error correction is shown in Fig.~\ref{circuit2EC}.
Since the QFI is invariant under unitary transformations (that are independent of $\theta$), we have omitted the decoding unitary $V$ in Fig.~\ref{circuit2EC}. 
Interestingly enough, the numerical results for error correction match those of error filtration. 

The reason can be seen by considering the optimal code in Eqs.~(\ref{encoding0_QFI})-(\ref{encoding1_QFI}).
Independent depolarizing noise maps the state (\ref{gfd3h1}) into

\begin{align}\label{chi1}
\chi_\theta 
& = q_r^2 |\Psi_\theta\rangle\langle \Psi_\theta | \nonumber \\
& \phantom{=}~ + 2q_r(1-q_r) \left[ 
\frac{\mathbb{I}}{4} + \frac{\sin{\theta}}{4} \left( |\psi_0\rangle \langle \psi_1| 
+ 
|\psi_1\rangle \langle \psi_0|
\right)
\right] \nonumber 
\\
& \phantom{=}~
+ (1-q_r^2) \frac{\mathbb{I}}{4} \, ,
\end{align}
where $\mathbb{I}$ is the identity matrix for two qubits.
We can rewrite $\chi_\theta$ as follows:
\begin{align}\label{chi2}
\chi_\theta 
& = q_r^2 |\Psi_\theta\rangle\langle \Psi_\theta | \nonumber \\
& \phantom{=}~ + 2q_r(1-q_r) \left[ 
\frac{\Pi}{4} + \frac{\sin{\theta}}{4} \left( |\psi_0\rangle \langle \psi_1| 
+ 
|\psi_1\rangle \langle \psi_0|
\right)
\right] 
\nonumber 
\\
& \phantom{=}~
+ 2q_r(1-q_r)  
\frac{\Pi}{4}
\nonumber 
\\
& \phantom{=}~
+ \left[(1-q_r^2) + 2q_r(1-q_r)\right]\frac{\mathbb{I}-\Pi}{4} \, ,
\end{align}
where $\Pi$ is the projector into the code space spanned by $|\psi_0\rangle$ and $|\psi_1\rangle$.
Within the error correction scheme of Fig.~\ref{circuit2EC}, we need to evaluate the QFI of the state $\chi_\theta$ in Eq.~(\ref{chi2}).
On the other hand, for the error filtration scheme of Fig.~\ref{fig:QFI_true}, and according to Eq.~(\ref{tauQFI}), the QFI must be evaluated on the state 
\begin{align}\label{chi3}
\tilde{\chi}_\theta 
& = q_r^2 |\Psi_\theta\rangle\langle \Psi_\theta | \nonumber \\
& \phantom{=}~ + 2q_r(1-q_r) \left[ 
\frac{\Pi}{4} + \frac{\sin{\theta}}{4} \left( |\psi_0\rangle \langle \psi_1| 
+ 
|\psi_1\rangle \langle \psi_0|
\right)
\right] 
\nonumber 
\\
& \phantom{=}~
+ 2q_r(1-q_r)  
\frac{\Pi}{4}
\nonumber 
\\
& \phantom{=}~
+ \left[(1-q_r^2) + 2q_r(1-q_r)\right]
|\omega \rangle \langle \omega| \, .
\end{align}
It is immediate to verify that 
the states $\chi_\theta$ and $\tilde{\chi}_\theta$ yield the same value for the QFI.

\begin{table*}[t]
\caption{The Table compares our results obtained with the error filtration scheme with other protocols designed for the error correction scheme where QFI is fully preserved \cite{lu2015robust} and Approximate Quantum Error-Correcting Code AQECC in Ref.~\cite{lin2024covariant}.  
An AQECC is a code that does not exactly correct all errors, but approximately preserves the encoded information under noise. Instead of requiring $\mathcal{R}\circ \mathcal{N} (\rho)=\rho$ (as in perfect QEC), it allows $||\mathcal{R}\circ \mathcal{N} (\rho)-\rho||\le \epsilon$, for small $\epsilon$.}
\label{tab:qfi}
\begin{ruledtabular}
\begin{tabular}{p{3.5cm} p{4.3cm} p{4.3cm} p{4.3cm}}
\textbf{Feature} & \textbf{Lu--Yu--Oh (2015)~\cite{lu2015robust}} & \textbf{Lin et al.\ (2024)~\cite{lin2024covariant}} & \textbf{Ali et al.\ (2025)} \\
Primary idea &
Preserve QFI directly; derive testable conditions (Eq.~4) and stabilizer-based constructions &
Use covariant approximate QECCs; bound code inaccuracy $\epsilon$ and show QFI loss $\le 4\epsilon M^2$ (Eq.~6, Prop.~27) &
Optimize short-depth error-filtration unitary $U$ with few ancillas \\
Resources &
$n = 2t+1$ physical qubits protect against $t$ arbitrary qubit errors after sensing &
$N$ spin-$s$ qudits; logical subspace labelled by $(J,M)$; logical dimension $k \approx (b-c)\log_2 N$ &
1 signal + $n \le 3$ ancillas; universal $U \in SU(2^{n+1})$ \\
Noise model handled &
Any channel whose Kraus ops lie in the span of a correctable set (digitalises errors) &
(i) Generic $d$-local noise $\Rightarrow \epsilon(d) = \mathcal{O}\!\big(\sqrt{d}(2s+1)^d N^{(b-a)/2}\big)$; (ii) Heralded erasure $\Rightarrow \epsilon = \mathcal{O}\!\big(N^{b + c/2 - 1}\big)$ &
Independent dephasing / depolarising on all qubits (signal and ancillas) \\
QFI after noise &
Exactly preserved if conditions satisfied; Heisenberg scaling retained for logical-GHZ &
$Q \ge (1 - 4\epsilon)\, 4M^2$; with $b > 1/2$ and $\epsilon \to 0$, one retains $Q \sim N^{2b}$ &
Closed-form QFI for $n=1$ ancilla (Eq.~\ref{eq:Q1}): $Q_1(q_r) = \dfrac{2q_r^2}{q_r^2 + 1}$, which is greater than $Q_0 = q_r^2$ without filtration \\
Recovery / measurement &
No recovery needed; optimal POVM = eigenbasis of common SLD &
Recovery optional; bound holds already for encoding-only (data-processing theorem) &
Decode ancillas (or keep them) before estimating $\theta$; no QEC cycle used \\
\end{tabular}
\end{ruledtabular}
\end{table*}


\section{Comparison with Superposed Quantum Error Mitigation (SQEM)}\label{sec:sqem}

\begin{figure}[tbh]
    \centering
    \includegraphics[width=0.9\linewidth]{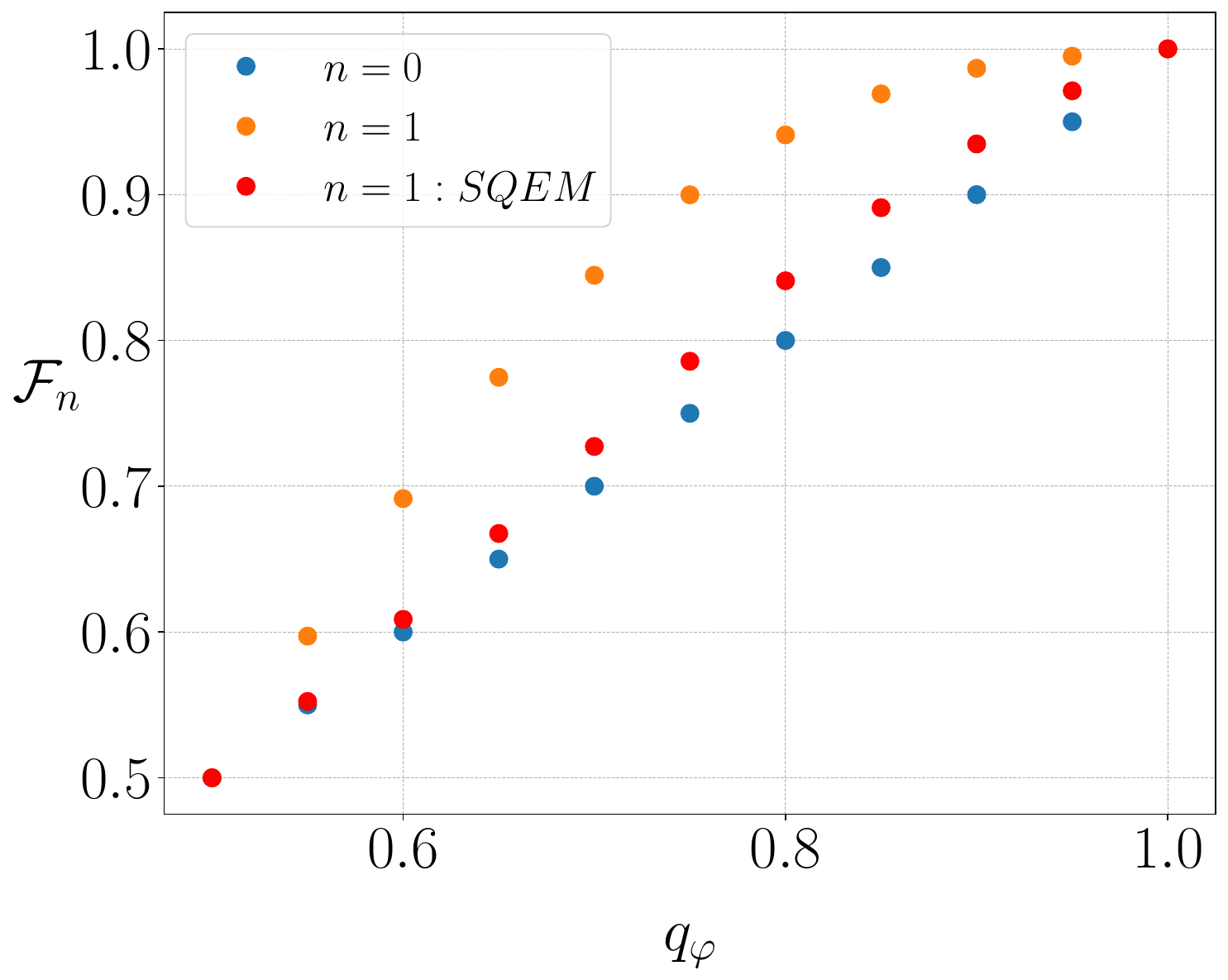}
    \caption{Comparison of entanglement fidelity $\mathcal{F}_n$ versus the dephasing noise parameter $q_\varphi$ for different approaches: the baseline circuit ($n=0$), our optimized error filtration protocol using one ancillary qubit, and the SQEM error mitigation scheme.}
    \label{fig:F_COMP}
\end{figure}

In this Section we compare our proposed optimized error filtration protocol to the recently introduced Superposed Quantum Error Mitigation (SQEM)~\cite{PhysRevLett.131.230601}.
In Fig.~\ref{fig:F_COMP}, we compare the entanglement fidelity $\mathcal{F}_n$ versus the dephasing noise parameter $q_\varphi$ for three scenarios. The baseline case ($n=0$) represents the circuit without any error mitigation. The second scenario employs our error filtration protocol using a single ancilla ($n=1$), while the third implements the SQEM protocol with one ancilla. In the figure, red markers denote SQEM results and orange markers represent our single-ancilla protocol. Notably, our approach consistently yields higher fidelity across the full noise range, surpassing SQEM at every tested value of $q_\varphi$.

These findings demonstrate that, under equivalent dephasing conditions, our error filtration scheme offers more robust protection of quantum information than SQEM. Overall, incorporating a single ancillary qubit as proposed significantly enhances fidelity.

\section{Conclusions}\label{sec:end}

We have analyzed the performance of optimized noise filtration protocols to enhance the reliability and accuracy of quantum operations, where large-scale fault-tolerant architectures are not yet feasible~\cite{IBMErrorHandling2022}. 
Our optimized scheme provides a useful tool in the portfolio of techniques~\cite{PhysRevA.72.012338,PhysRevLett.94.230501}, giving users the flexibility to choose the best option based on their accuracy needs and the amount of overhead they are willing to accept.
\textit{(i)} Our scheme yields non-trivial error filtration even with just one ancillary qubit. By contrast, previous schemes needed at least two assisting qubits to mitigate errors on a single qubit.
\textit{(ii)} We have studied a more realistic scenario in comparison to previous works~\cite{PhysRevLett.131.230601,PhysRevA.108.062604}, where the error reduction scheme was not robust to errors affecting the ancillary qubits. 
{\textit{(iii)} Filtration schemes can be obtained from error detection codes, and optimal scheme are seen to satisfy a particular \textit{error filtration condition}}.

We have examined scenarios where the encoding steps themselves are imperfect, introducing noisy ancillas and cross-talk. Even under these conditions, our protocol shows resilience, with a clear advantage as the number of ancillary qubits increases. In our application to metrology, the QFI of the optimal post-selected state equals that of the optimal non-post-selected state, i.e.~error filtration offers no advantage over deterministic error correction. It remains to understand whether this is a general equivalence or a specific feature of the model we have considered.
We have compared our proposal with the Superposed Quantum Error Mitigation (SQEM)~\cite{PhysRevLett.131.230601,PhysRevA.108.062604}, finding that our method provides consistently higher entanglement fidelities over a broad range of noise strengths. 
This comparative study underlines the robustness of our scheme, making it a promising candidate for near-term quantum devices.

Showing the success of our implementation is the first stage to propose experimental robust error reduction strategies, even in the presence of noise and cross-talk in the pre- and post-processing operations.
\textit{(i)} Though the current approach is shown to be robust under certain realistic models, the robustness analysis may be extended to more general noise models and put in relation to a suitable notion of fault-tolerance.
\textit{(ii)} Applications with small number of parameters are promising, yet real-time hardware platforms require a dynamic analysis to optimize parameters rapidly~\cite{renner2008security,kelly2015state}.
\textit{(iii)} Combining our method with a quantum channel followed by its quasi-inversion could also be of significant interest~\cite{Shahbeigi_2021}. 
\textit{(iv)} Future work may focus on testing the robustness of the method in multipartite Bell scenario for nonlocal quantum network~\cite{Paneru2024,Karczewski2022,Boreiri2025}.

\begin{acknowledgments}

We thank D. Pomarico, for introducing us HTCondor, and D. Amaro-Alcal\'a, F. Shahbeigi, R. Demkowicz-Dobra\'nski, and E. Galv\~ao for their valuable scientific support.
We are also grateful to S. P. Walborn,  G. Lima and their experimental quantum optics group of Universidad de Concepti\'on for fruitful discussions.
Finally, we warmly thank the anonymous referees for their valuable reviews.
This work has received funding from the 
the European Union's Horizon Europe research and innovation programme under the Project ``Quantum Secure Networks Partnership'' (QSNP, Grant Agreement No.~101114043);

European Union — Next Generation EU,
through
PNRR MUR project CN00000013 `Italian National Centre on HPC, Big Data and Quantum Computing',
PNRR MUR project PE0000023-NQSTI `National Quantum Science and Technology Institute', 
and
INFN through the project `QUANTUM'.
\end{acknowledgments}

\bibliography{chat-papers}

\end{document}